\documentclass[print,onecolumn,showpacs]{revtex4}
\usepackage{amsmath}
\usepackage{color,soul}
\usepackage{graphicx}
\usepackage{dcolumn}
\usepackage{bm}
\usepackage[dvipdfm,pdfstartview=FitH]{hyperref}

\begin{document}

\title{\textbf{The most generalized analytical approximation to the solution of single-mode spin-boson model
without rotating-wave approximation}}
\author{T. Liu$^{1,2}$, K. L. Wang$^{3}$, and M. Feng$^{1}$}
\email{mangfeng@wipm.ac.cn}
\affiliation{$^{1}$ State Key Laboratory of Magnetic Resonance and Atomic and Molecular
Physics, Wuhan Institute of Physics and Mathematics, Chinese Academy of
Sciences, Wuhan, 430071, China\\
$^{2}$ The School of Science, Southwest University of Science and
Technology, Mianyang 621010, China \\
$^{3}$ The Department of Modern Physics, University of Science and
Technology of China, Hefei 230026, China}
\pacs{42.50.-p, 03.65.Ge}

\begin{abstract}
The single-mode spin-boson model (SMSBM) has extensive application in different subfields of physics.
In the absence of rotating-wave approximation (RWA), we try to solve SMSBM analytically. We argue
that the analytical expression obtained is the most exact approximation to the solution of the
system under the assumption of Abel-Ruffini theorem, which works for a wide range of the parameters
such as coupling strength and detuning and would be practical for currently available
experiments.
\end{abstract}

\maketitle

The single-mode spin-boson model (SMSBM), describing a two-level
system experiencing a single-mode boson field, is an important
prototype in diverse phenomena in almost every subfield of physics.
From early day's studies of Holstein model \cite{holstein} in
condensed matter physics and Jaynes-Cummings model \cite{jc} in
quantum optics to recent investigation of quantum information
processing \cite{qip,qip1}, SMSBM has been playing crucial roles.

Rotating-wave approximation (RWA) has been usually employed in
treating SMSBM to simplify the solution under the condition of near
resonance and weak coupling within the characteristic time of the
system. However, RWA is not working well any more recently due to experimental
availability of strong coupling in atomic
system \cite{kimble}, semiconducting system \cite{semiconduct} and
superconducting system \cite{superconducting}, which implies the
necessity of solution to SMSBM in the absence of RWA.

Without RWA, however, it is hard to have an analytical solution to SMSBM
\cite{review} because the counter-rotating terms make the computational
subspace unclosed. As a result, no matter what methods were taken \cite%
{kus,reik1,reik,crisp,phoenix,twolevel,zaheer}, numerically or
semi-analytically, the solutions were made based on the truncated
subspace under some special conditions. Alternatively, as the
coherent state consists of infinite numbers of Fock states, the
computational subspace in the absence of RWA, which is unclosed in
the basis of Fock states, could be nearly closed in the
coherent-state representation for a wide parameter range. Therefore,
employment of coherent states would perhaps enable us to have
analytical approximations very close to the exact solution of the
problem.

There have been some peculiar characters discovered in SMSBM without
RWA, such as Bloch-Siegert shift \cite {bloch}, i.e., a shift with
respect to the true resonance frequency due to counter-rotating
terms, and quantum chaos in cavity QED by differently polarized
lights \cite {crisp}. It has also been shown that the discrepancy of
the SMSBM in the presence of RWA with respect to the absence of RWA
is reflected by some phase dependent effects \cite {phoenix}. Most
of those discrepancies are meaningful only theoretically, whereas
Bloch-Siegert shift is observable experimentally.  For example, in
ion traps, the strong laser radiation on the trapped ultracold ions
could lead to some level deviation regarding Bloch-Siegert shift
\cite {feng, liu, muga}. As the laser is usually treated as external
classical light, counter-rotating terms could show observable
effects in blue detuning case. Recent study has demonstrated the
possibility of fast logic gating with strong coupling between laser
and trapped-ion qubits, in which the usual RWA treatment could not
work well \cite {fast}, and the strong nonlinearity would yield
considerable complexity in the time evolution \cite {zheng, zeng, wang}.
As a result, it is of importance to have a more strict study of the
SMSBM without RWA.

We have noticed a very recent publication for analytically solving SMSBM by generalizing
RWA \cite {irish}. In the present paper, we will try to explore a more efficient
approach to a more generalized analytical result, compared to \cite {irish}.
The key point is that, by employing coherent-state
representation, we try to diagonilize the matrix regarding the
computational subspace. Based on the Abel-Ruffini theorem that no
general solution in radicals is possible to polynomial equations of
degree five or higher \cite {art}, we present analytically the expressions of
the energy levels of the system under third-order approximation of
the exact wavefunction, which is relevant to a polynomial equation
of degree four. We argue that this should be the most generalized approximation to
the solution of the SMSBM. To further show the validity of our
solution, we will give evidences both analytically and numerically
in comparison with the results in previous publications. Moreover, we will
show potential application of our result.

Consider following Hamiltonian in units of $\hbar =1$ \cite{irish},
\begin{equation}
H=\left(
\begin{array}{cc}
\omega _{0}a^{+}a+\lambda (a^{+}+a) & \Omega /2 \\
\Omega /2 & \omega _{0}a^{+}a-\lambda (a^{+}+a)%
\end{array}%
\right) ,  \label{1}
\end{equation}%
where $\Omega$ is the energy splitting of the spin and $\omega_{0}$
is the frequency of the boson field. $\lambda $ denotes the coupling
between the spin and the boson field, and $a$ $(a^{+})$ is the
annihilation (creation) operator of the boson field. Eq. (1) is taken from
\cite {irish} in order for us to make comparison between our solution and in
\cite {irish}. In fact, with respect to the standard quantum optical notation,
Eq. (1) has been taken a unitary rotation on the two-level system. However,
as there is no change of physical essence with that unitary transformation,
we will work on Eq. (1) in most of this paper.

For convenience of treatment, we may set $g=\lambda /\omega _{0}$
and employ
displacement operator $\hat{D}(g)=\exp {[g(a^{\dagger }-a)]}$ acting on $%
a^{\dagger }$ and $a$, which yields $A=\hat{D}(g)^{\dagger }a\hat{D}(g)=a+g$%
, $A^{\dagger }=\hat{D}(g)^{\dagger }a^{\dagger }\hat{D}(g)=a^{\dagger }+g$,
$B=\hat{D}(-g)^{\dagger }a\hat{D}(-g)=a-g$, and $B^{\dagger }=\hat{D}%
(-g)^{\dagger }a^{\dagger }\hat{D}(-g)=a^{\dagger }-g$. Consequently, Eq.
(1) could be rewritten as%
\begin{equation}
H_{1}=\left(
\begin{array}{cc}
\omega _{0}(A^{+}A-g^{2}) & \Omega /2 \\
\Omega /2 & \omega _{0}(B^{+}B-g^{2})%
\end{array}%
\right) ,
\end{equation}%
which is formally solvable, and we assume following trial solution to Eq. (2),%
\begin{equation}
|\rangle =\sum_{n=0}^{N}(c_{n}|n\rangle _{A}|e\rangle +d_{n}|n\rangle
_{B}|g\rangle ),
\end{equation}%
where $|e\rangle =\binom{1}{0}$ and $|g\rangle =\binom{0}{1}$, $c_{n}$ and $%
d_{n}$ are coefficients determined later, and $N$ is a large integer
relevant to the size of the truncated subspaces. $|n\rangle _{A(B)}$ is a coherent
state regarding the operator $A(B),$ which defines as
$|n\rangle _{A}=\frac{1}{\sqrt{n!}}(a^{\dagger }+g)^{n}|0\rangle _{A}$
and $|n\rangle _{B}=\frac{1}{\sqrt{n!}}(a^{\dagger }-g)^{n}|0\rangle _{B}$
with $|0\rangle_{A(B)}$ the coherent state in the subspace regarding the operator $A(B)$ \cite {liu}.

Putting Eq. (3) into the Schr\H{o}dinger equation of Eq. (2) could yield%
\begin{equation}
\omega _{0}(m-g^{2})c_{m}+\frac{\Omega }{2}\sum%
\limits_{n=0}^{N}(-1)^{n}D_{mn}d_{n}=Ec_{m},
\end{equation}%
\begin{equation}
\omega _{0}(m-g^{2})d_{m}+\frac{\Omega }{2}\sum%
\limits_{n=0}^{N}(-1)^{m}D_{mn}c_{n}=Ed_{m},
\end{equation}%
where $(-1)^{n}D_{mn}=$ $_{A}\langle m|n\rangle _{B},$ $(-1)^{m}D_{mn}=$ $%
_{B}\langle m|n\rangle _{A},$ and $D_{mn}=e^{-2g^{2}}\sum_{k=0}^{\min
[m,n]}(-1)^{-k}\frac{\sqrt{m!n!}(2g)^{m+n-2k}}{(m-k)!(n-k)!k!}$ \cite{liu}.
Eqs. (4) and (5) present the possibility to have a closed solution to the
problem. To analytically solve Eqs. (4) and (5), we set $d_{n}=\pm(-1)^{n}c_{n}$,
which yields $\omega_{0}(m-g^{2})c_{m}\pm \frac {\Omega}{2}\sum_{n}^{N}
 D_{mn}c_{n}=E^{\pm}c_{m}$. The eigen solution of the equation relies on following determinant,
\begin{equation}
\left|
\begin{array}{ccccc}
e^{\pm}_{0} & \Omega^{\pm}_{0,1} & \Omega^{\pm}_{0,2} & \cdots & \Omega^{\pm}_{0,N}  \\
\Omega^{\pm}_{1,0} & e^{\pm}_{1} & \Omega^{\pm}_{1,2} & \cdots & \Omega^{\pm}_{1,N}  \\
\Omega^{\pm}_{2,0} & \Omega^{\pm}_{2,1} & e^{\pm}_{2} & \cdots & \Omega^{\pm}_{2,N}  \\
\cdots & \cdots & \cdots & \cdots & \cdots \\
\Omega^{\pm}_{N,0} & \Omega^{\pm}_{N,1} & \Omega^{\pm}_{N,2} & \cdots & e^{\pm}_{N} \\
\end{array}
\right| =0,
\end{equation}
where
$e^{\pm}_{m}=\omega_{0}(m-g^{2})+\Omega^{\pm}_{m,m}-E^{\pm}=\epsilon^{\pm}_{m}-E^{\pm}$, with
$\Omega^{\pm}_{m,n}=\pm(1/2)\Omega
D_{mn}$. As the superscripts $\pm$ are consistent for the relevant
variables, the cases regarding superscripts $'+'$ and $'-'$ will be
treated independently. In principle, if we consider a large enough
value of $N$, Eq. (6) would lead us to a nearly exact solution to Eqs. (4)
and (5). However, in terms of Abel-Ruffini theorem, to have the analytical
results to the best, we have to reduce the determinant to
\begin{equation}
\left|
\begin{array}{ccccc}
e^{\pm}_{m} & \Omega^{\pm}_{m,m+1} & \Omega^{\pm}_{m,m+2} & \Omega^{\pm}_{m,m+3}  \\
\Omega^{\pm}_{m+1,m} & e^{\pm}_{m+1} & \Omega^{\pm}_{m+1,m+2} & \Omega^{\pm}_{m+1,m+3}  \\
\Omega^{\pm}_{m+2,m} & \Omega^{\pm}_{m+2,m+1} & e^{\pm}_{m+2} & \Omega^{\pm}_{m+2,m+3}  \\
\Omega^{\pm}_{m+3,m} & \Omega^{\pm}_{m+3,m+1} & \Omega^{\pm}_{m+3,m+2} & e^{\pm}_{m+3} \\
\end{array}
\right| =0,
\end{equation}
which leads to a polynomial equation with degree four and should be the most generalized
analytical approximation of the solution to the SMSBM under our consideration. Straightforward deduction to Eq. (7)
yields
\begin{eqnarray}
\lefteqn{E_{m}=-\frac
{1}{4}(\delta^{-}_{m}+\sqrt{8\chi^{-}_{m}+\delta^{-
2}_{m}-4\alpha^{-}_{m}})} \notag \\
& &{}-\frac {1}{4}\sqrt{(\delta^{-}_{m}+\sqrt{8\chi^{-}_{}+\delta^{-
2}_{m}-4\alpha^{-}_{m}})^{2}-16(\chi^{-}_{m}+ \frac
{\delta^{-}_{m}\chi^{-}_{m}-\beta^{-}_{m}}{\sqrt{8\chi^{-}_{m}+\delta^{-
2}_{m}-4\alpha^{-}_{m}}})},
\end{eqnarray}
with $m=0, 1$ corresponding to the ground and the first excited
states, respectively. Other excited states are
\begin{eqnarray}
\lefteqn{E_{m+2}^{\pm}=-\frac
{1}{4}(\delta^{\pm}_{m}+\sqrt{8\chi^{\pm}_{m}+\delta^{\pm
2}_{m}-4\alpha^{\pm}_{m}})} \notag \\
& &{} + \frac
{1}{4}\sqrt{(\delta^{\pm}_{m}+\sqrt{8\chi^{\pm}_{m}+\delta^{\pm
2}_{m}-4\alpha^{\pm}_{m}})^{2}-16(\chi^{\pm}_{m}+ \frac
{\delta^{\pm}_{m}\chi^{\pm}_{m}-\beta^{\pm}_{m}}{\sqrt{8\chi^{\pm}_{m}+\delta^{\pm
2}_{m}-4\alpha^{\pm}_{m}}})},
\end{eqnarray}
with $m= 0, 1, 2, \cdots$.
$\chi^{\pm}_{m}=(-q^{\pm}_{m}/2+\sqrt{q^{\pm 2}_{m}/4+p^{\pm
3}_{m}/27})^{1/3}+ (-q^{\pm}_{m}/2-\sqrt{q^{\pm 2}_{m}/4+p^{\pm
3}_{m}/27})^{1/3}+\alpha^{\pm}_{m}/6$, $p^{\pm}_{m}=-\alpha^{\pm
2}_{m}/12+\delta^{\pm}_{m}\beta^{\pm}_{m}/4-\gamma^{\pm}_{m}$,
$q^{\pm}_{m}=-\alpha^{\pm
3}_{m}/108+\delta^{\pm}_{m}\alpha^{\pm}_{m}\beta^{\pm}_{m}/24+
\alpha^{\pm}_{m}\gamma^{\pm}_{m}/3-\delta^{\pm
2}_{m}\gamma^{\pm}_{m}/8-\beta^{\pm 2}_{m}/8$,
$\delta^{\pm}_{m}=-\sum_{j=0}^{3}\epsilon^{\pm}_{m+j}$,
$\alpha^{\pm}_{m}=\sum_{j=1}^{3}(\epsilon^{\pm}_{m}\epsilon^{\pm}_{m+j}+\epsilon^{\pm}_{m+j}\epsilon^{\pm}_{m+j+1}
-\Omega^{\pm 2}_{m,m+j}-\Omega^{\pm 2}_{m+j,m+j+1})$,
$\beta^{\pm}_{m}=
-\sum_{j=0}^{3}[2\prod_{k=1}^{3}\Omega^{\pm}_{m+j+k,m+j+k+1}+
\epsilon^{\pm}_{m+j}(\epsilon^{\pm}_{m+j+1}\epsilon^{\pm}_{m+j+2}-
\sum_{k=1}^{3}\Omega^{\pm 2}_{m+j+k,m+j+k+1})]$, and
$$ \gamma^{\pm}_{m}=\left|
\begin{array}{ccccc}
\epsilon^{\pm}_{m} & \Omega^{\pm}_{m,m+1} & \Omega^{\pm}_{m,m+2} & \Omega^{\pm}_{m,m+3}  \\
\Omega^{\pm}_{m+1,m} & \epsilon^{\pm}_{m+1} & \Omega^{\pm}_{m+1,m+2} & \Omega^{\pm}_{m+1,m+3}  \\
\Omega^{\pm}_{m+2,m} & \Omega^{\pm}_{m+2,m+1} & \epsilon^{\pm}_{m+2} & \Omega^{\pm}_{m+2,m+3}  \\
\Omega^{\pm}_{m+3,m} & \Omega^{\pm}_{m+3,m+1} & \Omega^{\pm}_{m+3,m+2} & \epsilon^{\pm}_{m+3} \\
\end{array}
\right|.$$

To prove the validity of Eqs. (8) and (9) explicitly,
we may work along following two aspects: Comparison with numerical treatment of Eq.
(6) in the case of a big enough value of N, and comparison with other analytical
solutions by the determinants with smaller subspaces. For the former, we
have made numerics on Eq. (6), as shown in Fig. 1, by setting $N=42$
with off-diagonal elements $\Omega_{ij}$ $(i$ or $j \ge N$) smaller
than $10^{-6}$. We may consider that numerical result is the exact
solution to the problem. Fig. 1 shows that our results in Eqs. (8)
and (9) under the third-order approximation agree with the exact
solution very well even in the case that $\lambda$, $\Omega$ and $\omega_{0}$
are comparable.

\begin{figure}
\scalebox{0.7}{\includegraphics{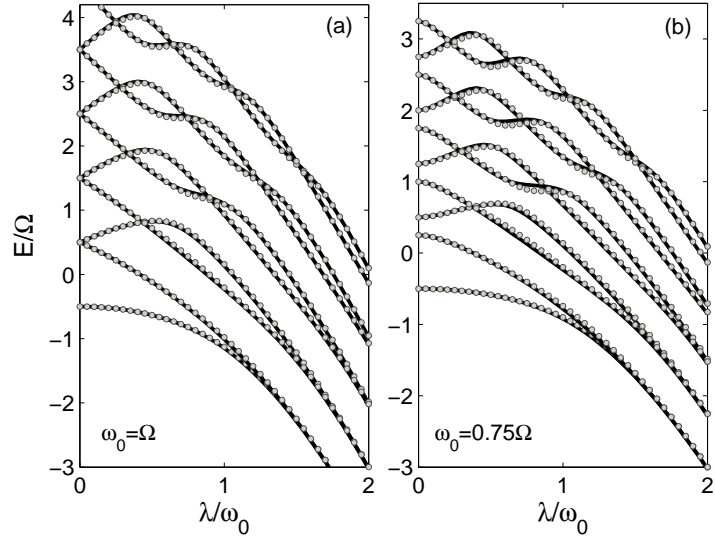}} \caption{Energy levels
with respect to ($\lambda/\omega_{0}$), where we assume (a)
$\omega_{0}=\Omega$ and (b) $\omega_{0}=0.75\Omega$. The solid curve
and circles, respectively, mean the numerical (exact) solution and
the solution from Eqs. (8) and (9) under the third-order
approximation.} \label{fig1}
\end{figure}

\begin{figure}
\scalebox{0.7}{\includegraphics{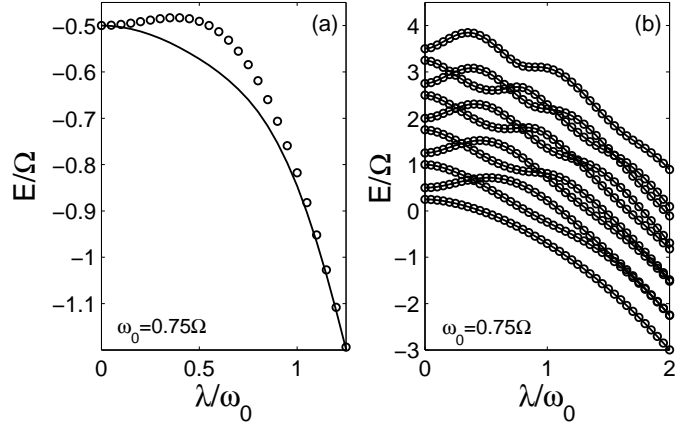}} \caption{Energy levels
with respect to ($\lambda/\omega_{0}$) by Eq. (10) under the
first-order approximation (solid curve) and by the results from
\cite {irish} (the circles), where we assume
$\omega_{0}=0.75\Omega$, (a) and (b) correspond to ground state and
excited states, respectively.} \label{fig2}
\end{figure}

For the latter, we first consider the zero-order approximation of
Eq. (6), i.e., $e_{m}=0$, which yields
$E^{\pm}_{m}=m\omega_{0}-(\lambda^{2}/\omega_0)\pm \Omega D_{mm}/2$.
The first-order approximation corresponds to
\begin{equation}
\left|
\begin{array}{cc}
e^{\pm}_{m} & \Omega^{\pm}_{m,m+1} \\
\Omega^{\pm}_{m+1,m} & e^{\pm}_{m+1}
\end{array}
\right| =0,
\end{equation}
which leads to the ground state energy,
$E_{0}=\omega_{0}(\frac{1}{2}-g^{2})-\frac
{\Omega}{4}(D_{0,0}+D_{1,1})-\frac {1}{2}\sqrt{[\omega_{0}+\frac
{\Omega}{2}(D_{0,0}-D_{1,1})]^{2} +\Omega^{2}D_{0,1}^{2}}$ and the
energies for excited states
$E^{\pm}_{k+1}=\omega_{0}(\frac{1}{2}+k-g^{2})+(-1)^{k}\frac
{\Omega}{4}(D_{k,k}+D_{k+1,k+1})\pm\frac
{1}{2}\sqrt{[\omega_{0}-(-1)^{k}\frac
{\Omega}{2}(D_{k,k}-D_{k+1,k+1})]^{2} +\Omega^{2}D_{k,k+1}^{2}}$,
where $k=0, 1, \cdots, N$ and $D_{k,k+1}=D_{k+1,k}$ is used. It
could be found in Fig. 2 by comparison with the results in \cite
{irish} that the excited-state energies we obtain are in good
agreement with those in \cite {irish}, but the ground state not.
In this context, we consider that the results obtained in \cite {irish}
is basically the one under the first-order approximation in our treatment.
But the ground state plotted in \cite
{irish} seems to be $E^{-}_{0}=-\Omega
D_{0,0}/2-\lambda^{2}/\omega_{0}$, i.e., the zeroth-order
approximation shown above \cite {question}.

Analogically, we may also obtain the second-order approximation
using
\begin{equation}
\left|
\begin{array}{ccc}
e^{\pm}_{m} & \Omega^{\pm}_{m,m+1} & \Omega^{\pm}_{m,m+2} \\
\Omega^{\pm}_{m+1,m} & e^{\pm}_{m+1} & \Omega^{\pm}_{m+1,m+2} \\
\Omega^{\pm}_{m+2,m} & \Omega^{\pm}_{m+2,m+1} & e^{\pm}_{m+2}\\
\end{array}
\right| =0,
\end{equation}
for which we omit the lengthy expression of the analytical result,
but emphasize that the accuracy of the solution depends on how many
off-diagonal terms in the determinant of Eq. (6) are involved. In
general, the farther the off-diagonal elements away from the
diagonal line of the matrix in Eq. (6), the less significant the
elements play their roles in the solution. But in the case that
$\Omega$, $\omega_{0}$ and $\lambda$ are comparable, our numerics
shows that the situation is very complicated, for example,
$\Omega_{3,0}$, $\Omega_{2,0}$, $\Omega_{0,3}$, $\Omega_{0,2}$
becoming comparable to $\Omega_{1,0}$ and $\Omega_{0,1}$. This is
the reason that omission of the elements other than the nearest
neighbor to the diagonal terms of the matrix yields the deviation in
the mediate coupling case in \cite {irish} with respect to the exact
solution. In contrast, our treatment could present results more
accurate than under the standard RWA and than in \cite {irish}. As demonstrated in Fig. 1,
our analytical expression fits the numerical results very well in a wide range of parameters.
Under the assumption of Abel-Ruffini theorem, we argue that the
results we present in Eqs. (8) and (9) under the third-order
approximation should be the most accurate analytical expression for the
energy levels of the SMSBM under our consideration.

Why could we make this ? The key reason is the correlation between
$c_{n}$ and $d_{n}$ we found, i.e., $d_{n}=\pm(-1)^{n}c_{n}$, in the
coherent-state representation. The coherent-state representation
helps us to have a close subspace for solution in the absence of
RWA, and the coefficient correlation significantly simplifies our
analytical deduction, which makes it available to reach the
expression under the third-order approximation.

Compared with purely numerical treatments, our analytical result
could present some physics more clearly. For example, for $n=0$ in
Eq. (3), we have the ground state
$|\rangle=(1/\sqrt{2})(|0\rangle_{A}|e\rangle-|0\rangle_{B}|g\rangle)$,
which implies that the ground state of the system always overlaps
with the upper level of the spin and would always keep evolving if
we involve the counter-rotating terms in our treatment. In contrast,
under the framework of RWA and even in generalized RWA treatment
\cite {irish}, the ground state of the system is always uncoupled
from other states and thereby remains unchanged no matter how strong
the interaction is. Other potential application could also be found
in \cite {liu}.

On the other hand, as our analytical result is very close to the
exact solution, we may employ it to study quantum behavior of the
SMSBM under arbitrary conditions. For example, we may accurately
calculate the dynamics of the system in the regimes of mediate and
strong coupling, which is helpful in experimentally exploring the
decoherence and operational infidelity regarding qubits in quantum
information science. Specifically, for strong coupling case in
trapped ion system, we may unitarily transform the original
Hamiltonian to a Hamiltonian very similar to Eq. (1) \cite {liu}.
$$H'=-\frac {\Omega_{0}}{2}\sigma_{x}+\nu a^{\dagger}a+g(a^{\dagger}+a)\sigma_{z}+\epsilon\sigma_{z}+g^{2},$$
where $\Omega_{0}$ is the Rabi frequency regarding laser-ion
coupling, $\nu$ is the trap frequency, $g$ is related to Lamb-Dicke
parameter, and $\epsilon$ is the detuning of the laser with respect
to the trapped ion. $\sigma_{x,z}$ are usual Pauli operators based
on the two levels of the ion \cite {liu}. As an example, we
demonstrate in Fig. 3 the time evolution of the population in the
lower level under the initial condition $\Omega_{0}/\nu=$ 1 or 3/4,
$\epsilon=0$ and $g=0.8$ with coherent state $\alpha=1.0$ in
vibration. As the results based on Eqs. (8) and (9) are nearly the
exact solution, we could study the dynamics efficiently from the
complicated evolution of the system. The figure also shows the
deviation of the results by \cite {irish} from the exact dynamics.
What is more, as $g=0.8$ implies the case beyond Lamb-Dicke limit
\cite {liu}, our analytical results are also useful for
understanding the behavior of the ions outside the Lamb-Dicke regime
\cite{steane} from a purely quantum mechanical viewpoint.

\begin{figure}
\scalebox{0.7}{\includegraphics{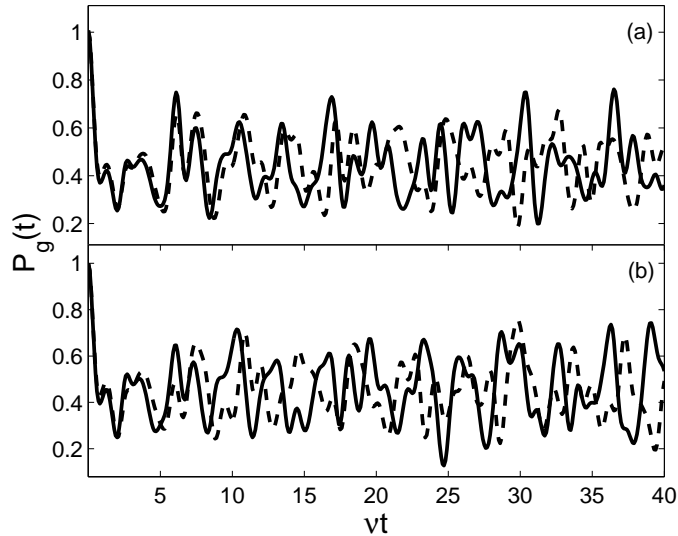}} \caption{Time
evolution of the population in the lower level of the trapped ion by
Eqs. (8) and (9) (solid curve) and by the first-order approximation
(dashed curve), where we assume $g=0.8$ and the initial vibrational
state in coherent state with $|\alpha|=1.0$. (a) $\Omega_{0}/\nu=1$
and (b) $\Omega_{0}/\nu=$3/4. We consider the solid curve is a nearly
exact description of the dynamics of the trapped ion. As the
generalized RWA treatment in \cite {irish} is equal to our treatment
under the first-order approximation, we may find the deviation of
the approximate treatment from the exact description with time.}
\label{fig3}
\end{figure}

It seems also possible to apply our method to more complicated situation
regarding strong spin-boson coupling. For example, extending SMSBM
to multi-spin single-mode boson interaction reaches Dicke model
\cite {dicke}. It has been found that the Dicke model in the absence
of RWA owns some unique characters \cite {chen} with respect to the
case in the presence of RWA \cite {buzek} such as in quantum phase
transition, in Berry phase and in entanglement. Moreover,
considering multi-mode boson field, we could have relevance to another
fundamental problem with environment interrupting a spin-qubit, as shown in a
recent work \cite {zhenghan} that Zeno effect is stronger in the absence of RWA than in the
presence of RWA, and anti-Zeno effect disappears if RWA is removed.
With our method by minor modification, we may enable some analytical
discussion for above relevant problems.

In conclusion, we have presented some analytical expressions for
solving the SMSBM in the absence of RWA. We argue that our
third-order approximation is the most accurate
analytical result, which could effectively replace the exact
numerical solution for studying the SMSBM under arbitrary condition.
As SMSBM has been widely applied to different physical problems
\cite {review,report}, the coherent-state forms of the eigenfunction
and the analytical expression of the energy levels of the system
would be helpful for understanding the interaction and the dynamics
in the spin-boson model under strong coupling or other extreme
conditions.

We also conjecture that our technique could be extended to
multi-spin or multi-mode cases. Compared to numerical solutions to
these cases, our analytical results would help us to get more
physical insight from the complexity.

The work is supported by NNSFC under Grant No. 10774163 and by the
funding from the State Key Lab of MRAMP.


\begin{thebibliography}{99}
\bibitem{question} We have compared repeatedly our calculations with the results in \cite {irish}
on this point. We conjecture that Ref. \cite {irish} confused the
results of zeroth- and first-order approximations in the treatment.
Although this need further clarification, as all the other energies
regarding the excited states are in good agreement, we consider it
as a good justification for our key result, i.e., Eqs. (9) and (10).
\end{thebibliography}
\end{document}